\documentclass[preprintnumbers,amsmath,amssymb,floatfix,10pt,prd,onecolumn,
superscriptaddress,nofootinbib]{revtex4}
\usepackage{latexsym,float}
\usepackage{epsfig,color}
\usepackage{epstopdf}
\usepackage{amssymb}
\usepackage{verbatim}
\usepackage{multirow}
\usepackage[utf8]{inputenc}

\begin{document}

\title{\bf Complexity analysis of Cylindrically Symmetric Self-gravitating Dynamical
System in $f(R,T)$ Theory of Gravity}

\author{M. Zubair}
\email{mzubairkk@gmail.com; drmzubair@cuilahore.edu.pk}\affiliation{Department of Mathematics,
COMSATS University Islamabad, Lahore Campus, Lahore-Pakistan}

\author{Hina Azmat}
\email{hinaazmat0959@gmail.com}\affiliation{Department of
Mathematics, COMSATS University Islamabad, Lahore Campus, Lahore-Pakistan}

\date{\today}

\begin{abstract}
In this article, we have studied a cylindrically symmetric self-gravitating dynamical
object via complexity factor which is obtained through orthogonal splitting of Reimann tensor in $f(R,T)$ theory of gravity. Our study is based on the definition of complexity for dynamical sources, proposed
by Herrera \cite{12b}. We actually want to analyze the behavior of complexity factor for cylindrically symmetric dynamical source in modified theory. For this, we define the scalar functions through orthogonal splitting of Reimann tensor in $f(R,T)$ gravity and work out structure scalars for cylindrical geometry. We evaluated the complexity of the structure and also analyzed the complexity of the evolutionary
patterns of the system under consideration. In order to present simplest mode of
evolution, we explored homologous condition and homogeneous expansion condition
in $f(R,T)$ gravity and discussed dynamics and kinematics in the background of a
generic viable non-minimally coupled $f(R,T)=\alpha_1 R^m T^n +\alpha_2 T(1+\alpha_3 T^p R^q)$ model.
In order to make a comprehensive analysis, we considered three different
cases (representing both minimal and non-minimal coupling) of the model under
consideration and found that complexity of a system is increased in the
presence of higher order curvature terms, even in the simplest modes of evolution.
However, higher order trace terms affects the complexity of the system but
they are not crucial for simplest modes of evolution in the case of minimal coupling. The stability of vanishing of complexity factor has also been discussed.\\

\end{abstract}

\maketitle

\section{Introduction}
The characterization of cosmological structures on the basis of their degree of complexity
is an interesting physical phenomenon as it is directly related to the complications
appearing in a stellar system due to its physical and structural characteristics.
The measure of degree of complexity can be helpful in the study of the formation and
evolution of stellar objects.All those factors that triggers the complications
in a system are also involved in the different evolutionary phases of a system.
However, a precise and concise definition of complexity in the field of astrophysics
is still an ambition and aspiration of the community of astrophysicists. In past, many
sincere efforts have been made in order to define an appropriate
criterion for the measure of degree of the complexity in different branches
of science \cite{1}-\cite{10}. Among the many definitions that have been proposed
so far, most of them are related to the concepts of information and disequilibrium.
These are based on the natural thought that complexity is related to a basic property
describing the structures existing within the system. In physics, perfect crystals and
ideal gas represent two different but simplest models, so they can be considered as
systems with zero complexity. Nevertheless, study of both of these systems ensure
that definition of complexity cannot be confined to the concepts of entropy and
information, rather it includes some other factors which was going to be ignored.

A definition of complexity for self-gravitating systems is introduced in \cite{11}
which is based on the work developed by Lopez-Ruiz and his collaborators \cite{6}.
In this definition, probability distribution which appear in the definition of entropy
and information is replaced by energy density of the
fluids, which has been justified by the argument that this physical quantity is
related to the probability of finding some particles at given specified location
inside the star, or it proved difficult to suggest a better alternative from the
available physical quantities. Nevertheless, this definition has some important
drawbacks, as it only encompasses the role of energy density whereas other
physical parameters like pressure isotropy or anisotropy which are expected
to play an important role in structure formation of a system are completely
ignored. In order to avoid these drawbacks, an entirely new definition of complexity \cite{12} has
been proposed for spherically symmetric and static self-gravitating source. It is
devised on the basic assumption that less complex systems correspond to the fluid
configurations with homogeneous energy density and isotropic pressure.
Such distributions are assigned with zero degree of complexity factor
which appears in orthogonal splitting of Riemann tensor. Herrera and his collaborators extended this concept
of complexity from static to non-static scenario, where they consider not only the complexity factor
of structure of fluid configuration but also discuss the conditions of minimum
complexity of evolutionary patterns \cite{12a}. For an axially symmetric static source,
they explored three different complexity factors and found that all the three factors vanish for simplest fluid configuration \cite{12b}.
The significance of this definition motivated the researchers to explore
it for different scenarios. Herrera et al. extended this fascinating concept to the vacuum solutions
with the help of the Bondi metric which covers a huge number of spacetimes like
Minkowski spacetime, the static Weyl
metrics, gravitationally
radiating metrics, non radiative and non static metrics \cite{LH}, while Casadio et al.
studied isotropization and change of complexity
employing gravitational decoupling approach for
a static and spherically symmetric system \cite{CAS}.
In \cite{13}, the effects of electric charge has been incorporated and complexity factor has been
analyzed. They have found that electromagnetic field has considerable impacts on the complexity
of a cosmological structure. The same authors discussed the complexity of cylindrically symmetric static source in \cite{14}.

Cosmological objects are usually studied with the consideration of
spherical symmetry because observational evidences show that deformations
in spherical symmetry are very rare, however non-spherical symmetries may also exit
and provide significant information about celestial objects. Different cosmological
issues have also been discussed by assuming non-spherical situations. Cylindrically
symmetric sources grasped the attention of the community of the relativists when
Levi-Civita found its vacuum solution. Cylindrically symmetric thin-shell
wormholes, cylindrical polytropes with generalized polytropic equation of
state, charged non-adiabatic and perfect fluid with cylindrically symmetric
background were studied in \cite{17}-\cite{21}. Cylindrically symmetric
self-gravitating objects with anisotropic background evolving under
different conditions \cite{22}-\cite{24} have been explored in order to
understand different phases of evolution. Herrera and his collaborators \cite{25}
developed structure scalars for cylindrically symmetric and studied
dissipative fluid distribution with the help of these scalars.

General Relativity (GR) is still one of the most comprehensive theory in order
to understand the dynamics of the universe in accordance with its matter components. However, some cosmological issues like
unification of gravitation and quantum mechanics covering the singularity
problem and late time accelerated expansion of the universe necessitated the
improvements in theoretical framework of GR through modified theories. A large
number of class of modified theories has been introduced to overcome the
cosmological issues. Due to their significance, different cosmological
phenomena like anisotropy, luminosity and stability analysis of celestial
objects have been discussed and presented in literature with the help of
these theories \cite{15,16,22,23,24}. Reverberi \cite{27} studied the contracting dust
cloud with the $f(R)$ model and found that increase in energy density
bring about the curvature singularity. Cembranos and his co-authors \cite{28}
explored the inflation candidate in the context of a spherically symmetric
self-gravitating collapsing dust cloud in $f(R)$ gravity. Gravitational collapse and stability constraints for spherical and axial
symmetry has been studied via $f(R,T)$ gravity in \cite{37}-\cite{39}.

Compact stars have also been examined and explored in details with the help of modified gravities.
Strong gravity regime in viable models of $f(R)$ gravity has been studied \cite{40} and the claim that
stars with relativistically deep potentials cannot exist in $f(R)$ gravity has been disproved \cite{41}.
Capozziello et al. \cite{42} considered modified Lane-Emden equation that comes out from $f(R)$ gravity and
discussed the hydrostatic equilibrium of stellar structures.
Some interior models of compact stars like $4U1820-30, Her X-1, SAX J1808-3658$
has been studied using Krori and Barua analytical solution to the static spacetime with fluid
source in modified $f(R)$ gravity \cite{43}. Polytropic stars in Palatini $f(R)$-theories has also been investigated and it was shown how findings rely on regularity of the function $f(R)$ \cite{44}. The possible formation of compact astrophysical objects and their physical features has been discussed in the framework of $f(R, T)$ and $f(G, T)$ gravity \cite{45}-\cite{49}. Abbas and Nazar \cite{15, 16} discussed the effects of $f(R)$ gravity on complexity factor for spherically symmetric static source with anisotropic background.
In \cite{s1, s2}, complexity factor has been explored for static anisotropic sources in the context of Brans-Dick theory, while in \cite{s3, s4} it has been examined for a dynamical sphere and a class of compact stars in the framework of $f(R, T)$ gravity. The physical properties of compact objects were analyzed and their numerical outputs were found for different values of coupling parameter. Keeping in view the significance of modified theories, we intended to explore the definition of complexity factor for a cylindrically symmetric dynamical source with anisotropic fluid distribution in $f(R,T)$ gravity.

This article has been organized as follows: Next section presents basic notations, variables and set of field equations for cylindrically symmetric self-gravitating dynamical source in $f(R,T)$ gravity. Section III and IV entail the construction of structure scalars for self-gravitating cylinder in $f(R,T)$ gravity and the identification of complexity factor, respectively. Section V covers the discussion regarding simplest modes of evolution, while the section VI comprises the $f(R,T)$ model which is followed by the discussion about kinematics and dynamics of the system for three different cases of $f(R,T)$ model. Section VII explores the
stability of vanishing complexity factor condition. Last section concludes our results.

\section{Basic notations, variables and relativistic system of Equations in $f(R,T)$ theory of gravity}

We have considered an anisotropic stellar configuration with cylindrical geometry
which experiences the dissipation in the form of heat flux and its
interior region is given by the following expression
\begin{equation}\label{1}
ds^2_-=-F^2(t,r)dt^{2}+G^2(t,r)dr^{2}+H^2(t,r)\left(d\theta^{2}+
dz^2\right),
\end{equation}
Here, $F$ and $G$ are dimensionless, while $H$ has the same dimension as $r$ has.

The $f(R,T)$ modification of Einstein-Hilbert action is given by \cite{1*}
\begin{equation}\label{2}
\int dx^4\sqrt{-g}\left[\frac{f(R, T)}{16\pi G}+\mathcal{L} _ {(m)}\right].
\end{equation}

Here, action due to matter is described by $\mathcal{L} _ {(m)}$,
whose different choices can be made, each choice represents a particular
form of fluid.
Varying the modified action given in Eq.$(\ref{2})$ with respect to metric $g_{\alpha\beta}$,
we have the following set of field equations
\begin{eqnarray}\label{4}
&&R_{\alpha\beta} f_R(R,T)-\frac{1}{2}g_{\alpha\beta} f(R,T)+(g_{\alpha\beta}\Box-\nabla_{\alpha}\nabla_{\beta})f_R(R,T)=8\pi GT_{\alpha\beta}^{(m)}-f_T(R,T)T_{\alpha\beta}^{(m)}-f_T(R,T)\Theta_{\alpha\beta},
\end{eqnarray}
where $f_R(R,T)=\frac{\partial f(R,T)}{\partial R}$, $f_T(R,T)=\frac{\partial f(R,T)}{\partial T}$,
while $\nabla_{\alpha}$  is the covariant derivative
and $\Box$ is four-dimensional Levi-Civita covariant derivative.
The term $\Theta_{\alpha\beta}$ has the following mathematical representation
\begin{eqnarray}\label{5}
\Theta_{\alpha\beta}=\frac{g^{\rho\nu}\delta T_{\mu\nu}}{\delta g^{\alpha\beta}}= -2T_{\alpha\beta}+g_{\alpha\beta}\mathcal{L} _ {(m)}-2g^{\rho\nu}\frac{\partial^2\mathcal{L} _ {(m)}}{\partial g^{\alpha\beta}\partial g^{\rho\nu}}.
\end{eqnarray}
With the choice of $\mathcal{L} _ {(m)}= \mu$ (energy density) and $8\pi G = 1$, $\Theta_{\alpha\beta}$ takes the form as
\begin{eqnarray}\label{6}
\Theta_{\alpha\beta}=-2T_{\alpha\beta}+\mu g_{\alpha\beta}.
\end{eqnarray}
Using Eq.$(\ref{5})$, the modified field equations given in (\ref{4}) become
\begin{eqnarray}\label{7}
G_{\alpha\beta}&=& T_{\alpha\beta}^{eff},
\end{eqnarray}
where
\begin{eqnarray}\label{8}
T^{eff}_{\alpha\beta}&=&\frac{1}{f_R}\left[(f_T+1)T^{(m)}_{\alpha\beta}-\mu g_{\alpha\beta}f_T+
\frac{f-Rf_R}{2}g_{\alpha\beta}+(\nabla_\alpha\nabla_\beta-g_{\alpha\beta}\Box)f_R\right],
\end{eqnarray}
where $T^{(m)}_{\alpha\beta}$ represents the energy momentum tensor for the usual
matter.

 We have taken into account a fluid distribution
which is locally anisotropic and suffering dissipation in the form of heat flux. Its energy momentum
tensor has the following mathematical expression
\begin{eqnarray}\label{9}
T^{(m)}_{\alpha\beta}=(\mu+P_\perp)V_{\alpha}V_{\beta}-P_\perp g_{\alpha\beta}
+(P_r-P_\perp)\chi_{\alpha}\chi_{\beta}+q_\alpha V_\beta+V_\alpha q_\beta,
\end{eqnarray}
where $P_r$ and $P_\perp$ are two principal stresses. In most general cylindrical case,
one has three principal stresses which leads to an anisotropic tensor depending on two independent
scalar functions \cite{25}. However, in our case it will lead to an anisotropic tensor depending on one scalar
function. It will ultimately lead to a single scalar dependent structure scalar, which can further be analyzed
for complexity factor. We are interested in
structure scalar depending on single scalar instead of two, thus our assumption for fluid distribution is based on
two unequal principal stresses.
Here $V_{\beta}$, $\chi_{\beta}$ and $q_\alpha$ denote four-velocity,
unit four-vector along radial direction and heat flux respectively.
Under co-moving relative motion, these quantities are defined as
\begin{equation}\label{10}
V_{\alpha}= F^{-1}\delta^{0}_{\alpha},\quad
\chi_{\alpha}=G^{-1}\delta^1_{\alpha},\quad
q^\alpha=qG^{-1}\delta_1^\alpha,
\end{equation}
and satisfy the following relations
\begin{eqnarray}\nonumber
  V^{\alpha}V_{\alpha}=-1,\quad
  \chi^{\alpha}\chi_{\alpha}=1,\quad
  \chi^{\alpha}V_\alpha=0,\quad
  V^\alpha q_\alpha=0.
\end{eqnarray}
Eq.$(\ref{9})$ can also be expressed as
\begin{eqnarray}\label{T}
  T_{\alpha\beta}^{(m)} &=& \mu V_\alpha V_\beta+P h_{\alpha\beta}+\Pi_{\alpha\beta}+q(V_\alpha\chi_\beta+\chi_\alpha V_\beta),
\end{eqnarray}
where
\begin{eqnarray}\label{T1}
 P&=&\frac{1}{3}\left(P_r+2P_\perp\right),\quad h_{\alpha\beta}=g_{\alpha\beta}+V_\alpha V_\beta, \quad \Pi_{\alpha\beta}
 =\Pi\left(\chi_\alpha\chi_\beta-\frac{1}{3}h_{\alpha\beta}\right), \quad \Pi=P_r-P_\perp.
\end{eqnarray}

while the components of shear tensor $\sigma_{\alpha\beta}$ are defined as
\begin{eqnarray}\label{11}
&&\sigma_{\alpha\beta}=V_{(\alpha;\beta)}-a_{(\alpha}V_{\beta)}-\frac{1}{3}\Theta\left(g_{\alpha\beta}-V_\alpha V _\beta\right),
\end{eqnarray}
Here $a_\alpha$ is four-acceleration and $\Theta$ is expansion scalar which defines the rate of infinitesimal change of matter distribution. These two quantities
are defined by the following mathematical formulae
\begin{equation}\label{12}
a_\alpha=V_{(\alpha;\beta)}V^\beta,\quad
\Theta=V^\alpha_{;\alpha}.
\end{equation}
Four acceleration $a_\alpha$, expansion scalar $\Theta$ and non- zero components of shear tensor $\sigma_{\alpha\beta}$ can easily be calculated for given fluid distribution of cylindrically symmetric
gravitational source and these are given below
\begin{eqnarray}\label{13}
a_1=\frac{F'}{F},\quad
\Theta=\frac{1}{F}\left(\frac{\dot{G}}{G}+\frac{2\dot{H}}{H}\right),
\\\label{gg}
   \sigma_{11}=\frac{2}{\sqrt{3}}G^2\sigma, \quad\quad
   \sigma_{22}=\sigma_{33}=-\frac{1}{\sqrt{3}}H^2\sigma
\end{eqnarray}
where
  \begin{eqnarray}\label{sh}
  \sigma&=&\frac{1}{\sqrt{3}F}\left(\frac{\dot{G}}{G}-\frac{\dot{H}}{H}\right),
 \end{eqnarray}
 and its scalar value takes the form as
 \begin{eqnarray}\label{mi}
   \sigma^{\alpha\beta}\sigma_{\alpha\beta} &=& 2\sigma^2.
 \end{eqnarray}

The set of $f(R,T)$ field equations for the given cylindrically symmetric interior metric is given by
\begin{eqnarray}\label{16}
G_{00}&=&\frac{A^{2}}{f_R}\left[\mu+\Psi+\psi_{00}\right],\\\label{17}
G_{01}&=&\frac{FG}{f_R}\left((1+f_T)(-q)+\frac{\psi_{01}}{FG}\right),\\\label{18}
G_{11}&=&\frac{G^{2}}{f_R}\left[(1+f_T)\left(P_r\right)+\mu f_T-\Psi+\psi_{11}\right],\\\label{19}
G_{22}&=&\frac{H^{2}}{f_R}\left[(1+f_T)\left(P_\perp\right)+\mu f_T-\Psi+\psi_{22}\right],\\\nonumber
\end{eqnarray}
where
\begin{eqnarray}\label{20}
\Psi&=&\frac{f-Rf_R}{2},\quad
\psi_{00}=\frac{f''_R}{G^2}-\frac{\dot{f_R}}{F^2}\left(\frac{\dot{G}}{G}+2\frac{\dot{H}}{H}\right)
+\frac{f_R'}{G^2}\left(2\frac{H'}{H}-\frac{G'}{G}\right),\\\label{21}
\psi_{01}&=&\dot{f'_R}-\frac{F'}{F}\dot{f_R}-\frac{\dot{G}}{G}f'_R,\\\label{22}
\psi_{11}
&=&\frac{\ddot{f_R}}{F^{2}}-\frac{\dot{f_R}}{F^{2}}\left(\frac{\dot{F}}{F}-2\frac{\dot{H}}{H}\right)
-\frac{f'_R}{B^{2}}\left(\frac{F'}{F}+2\frac{H'}{H}\right),\\\label{23}
\psi_{22}&=&\frac{\ddot{f_R}}{F^{2}}-\frac{f''_R}{G^{2}}-\frac{\dot{f_R}}{F^{2}}\left(\frac{\dot{F}}{F}-\frac{\dot{G}}{G}-\frac{\dot{H}}{H}\right)
-\frac{f'_R}{G^{2}}\left(\frac{F'}{F}-\frac{G'}{G}+\frac{H'}{H}\right).
\end{eqnarray}

Thorne \cite{s5} proposed the idea that total amount of energy in a
cylindrical celestial object can defined through the gravitational C-energy, which takes the following
form for the case under consideration
\begin{eqnarray}\label{24}
m(t,r)&=&\left\{\left(\frac{\dot{H}}{F}\right)^2-\left(\frac{H'}{G}\right)^2\right\}\frac{H}{2}+\frac{l}{8}.
\end{eqnarray}
 Before moving towards the further computations,
it is worthwhile to define some notations.
The $D_T$ and $D_H$ are operators which represent proper time and radial derivatives, respectively,
and are defined as
\begin{eqnarray}\label{25}
&&D_T=\frac{1}{F}\frac{\partial}{\partial t},\quad
D_H=\frac{1}{H'}\frac{\partial}{\partial r},
\end{eqnarray}
whereas relativistic velocity of interior of collapsing fluid is given by
\begin{eqnarray}\label{26}
&&U=D_TH=\frac{\dot{H}}{F}<0,
\end{eqnarray}
From Eq.$(\ref{24})$, we can obtain
\begin{eqnarray}\label{27}
&&\tilde{E}=\frac{H'}{G}=\sqrt{\frac{l}{4H}+U^2-\frac{2}{H}m(t,r)}.
\end{eqnarray}

Using above equation together with Eq.$(\ref{17})$, we can develop the expression given below
\begin{eqnarray}\label{28}
\tilde{E}\left(\sqrt{3}\frac{\sigma}{H}-\frac{1}{3}D_H(\Theta-\sqrt{3}\sigma)\right)   &=& \frac{1}{2 f_R}\left(-q(1+f_T)+\frac{\psi_{01}}{FG}\right).
\end{eqnarray}
Eq.$(\ref{24})$ together with $(\ref{16})-(\ref{19})$ and $(\ref{25})$ provides
\begin{eqnarray}\label{29}
D_Tm   &=&\frac{H^2}{2f_R}\left\{-(1+f_T)\tilde{E}q-(1+f_T)UP_r+f_T\mu U+\frac{\tilde{E}}{FG}\pi_{01}-U\left(\Psi+\psi_{11}\right)\right\},
\end{eqnarray}
whereas radial derivative of mass provides
\begin{eqnarray}\label{30}
  D_Hm &=& \frac{H^2}{2f_R}\left\{\mu+\frac{U}{\tilde{E}}(1+f_T)q+\pi+\pi_{00}-\frac{U}{\tilde{E}}\frac{\psi_{01}}{FG}\right\},
\end{eqnarray}
which further leads towards the following expression
\begin{eqnarray}\label{31}
  m &=&\frac{1}{2}\int_0^r \frac{H^2}{f_R}\left\{\mu+\frac{U}{\tilde{E}}(1+f_T)q+\psi+\psi_{00}-\frac{U}{\tilde{E}}\frac{\psi_{01}}{FG}\right\}H'dr,
\end{eqnarray}
It can also be written as
\begin{eqnarray}\label{32}
 \frac{3m}{H^3} &=&\frac{3}{2H^3}\int_0^r \frac{H^2}{f_R}\left\{\mu+\frac{U}{\tilde{E}}(1+f_T)q+\psi+\psi_{00}-\frac{U}{\tilde{E}}\frac{\psi_{01}}{FG}\right\}H'dr.
\end{eqnarray}
\section{Weyl tensor and Structure scalars}

In order to define the structures scalars, we first need to find the weyl tensor which has two parts, i.e,
electric and magnetic parts. The electric part
is given below
\begin{eqnarray}\label{w}
  E_{\alpha\beta} &=& C_{\alpha\mu\beta\nu}V^{\mu}V^{\nu},
\end{eqnarray}

The non-trivial components of electric component of weyl tensor are
\begin{eqnarray}\label{E}
  E_{11} = \frac{2}{3}G^2\eta,\quad  E_{22}= -\frac{1}{3}H^2\eta=E_{33},
\end{eqnarray}
where
\begin{eqnarray}\nonumber
 \eta &=& \frac{1}{2F^2}\left\{\frac{\ddot{H}}{H}-\frac{\ddot{G}}{G}-\left(\frac{\dot{H}}{H}
 -\frac{\dot{G}}{G}\right)\left(\frac{\dot{F}}{F}+\frac{\dot{H}}{H}\right)\right\}\\\label{ee}
&&+\frac{1}{2B^2}\left\{\frac{F''}{F}-\frac{H''}{H}+\left(\frac{G'}{G}+\frac{H'}{H}\right)
\left(\frac{H'}{H}-\frac{F'}{F}\right)\right\}-\frac{1}{2H^2}.
\end{eqnarray}
With the help of $(\ref{24})$ and $(\ref{32})$, we can find the expression for the above scalar value
\begin{eqnarray}\nonumber
  \eta &=& \frac{1}{2f_R}\left[\mu-(1+f_T)\Pi+\Psi+\psi_{00}-\psi_{11}+\psi_{22}\right]
  \\\label{ex}&&-\frac{3}{2H^3}\int_0^r \frac{H^2}{f_R}\left\{\mu+\frac{U}{E}(1+f_T)q-\Psi+\psi_{00}
  -\frac{U}{E}\frac{\psi_{01}}{FG}\right\}H'dr,
\end{eqnarray}
The electric component $E_{\alpha\beta}$, in view of unit four-velocity and four-vectors can be given by
\begin{eqnarray}
  E_{\alpha\beta} &=& \eta(\chi_\alpha\chi_\beta-\frac{1}{3}h_{\alpha\beta})
\end{eqnarray}

Following Bel \cite{Bel} and Herrera et al. \cite{H1}-\cite{HRv}, we develop formalism for structure scalars in $f(R,T)$ gravity
and introduce a couple of tensors, named $Y_{\alpha\beta}$ and $X_{\alpha\beta}$.
For this, we orthogonally decompose the Riemann curvature tensor and find that
\begin{eqnarray}\label{xx}
  X_{\alpha\beta}&=& \frac{1}{3f_R}\left[\mu+\Psi+\psi_{00}\right]h_{\alpha\beta}-\frac{1}{2f_R}\left[(1+f_T)\Pi+\psi_{11}-\psi_{22}\right]
  \left(\chi_\alpha\chi_\beta-\frac{1}{3}h_{\alpha\beta}\right)-E_{\alpha\beta},\\\nonumber
  Y_{\alpha\beta}&=& \frac{1}{6f_R}\left[\mu+3f_T\mu+(1+f_T)(3P_r-2\Pi)+\Psi+\psi_{00}+\psi_{11}+2\psi_{22}\right]h_{\alpha\beta}-\frac{1}{2f_R}\left[(1+f_T)\Pi
  \right.\\\label{yy}&&+\left.\psi_{11}-\psi_{22}\right]
  \left(\chi_\alpha\chi_\beta-\frac{1}{3}h_{\alpha\beta}\right)+E_{\alpha\beta}.
 \end{eqnarray}
For detailed discussion of these quantities, one can see \cite{HRv}.
These tensors can be written in the combination of  structure scalars ($X_T$, $X_{TF}$, $Y_T$ and $Y_{TF}$).
\begin{eqnarray}\label{X}
  X_{\alpha\beta} &=& \frac{1}{3}X_T h_{\alpha\beta}+X_{TF}\left(\chi_\alpha \chi_\beta-\frac{1}{3}h_{\alpha\beta}\right),\\\label{Y}
   Y_{\alpha\beta} &=& \frac{1}{3}Y_T h_{\alpha\beta}+Y_{TF}\left(\chi_\alpha \chi_\beta-\frac{1}{3}h_{\alpha\beta}\right).
\end{eqnarray}
By making use of Eqs.$(\ref{16})$, $(\ref{18})$, $(\ref{19})$, $(\ref{24})$ and $(\ref{ex})$, we have the following expression
\begin{eqnarray}\label{Y1}
  \frac{3}{H^3}\left(m-\frac{l}{8}\right)&=& \frac{1}{2f_R}\left(\mu+\psi-(1+f_T)\Pi+\psi_{00}-\pi_{11}+\psi_{22}\right)-\eta,
\end{eqnarray}
which makes it possible to produce the following expression with the help of Eqs.(\ref{32}) and (\ref{Y1})
\begin{eqnarray}\nonumber
  Y_{TF} &=& \frac{1}{2f_R}\left\{\mu-2(1+f_T)\Pi+\pi+\pi_{00}-2\psi_{11}+2\pi_{22}\right\}-\frac{3}{2H^3}\int_0^r\frac{H^2}{f_R}\left\{
  \mu+\psi+\psi_{00}\right.\\\label{Y}&&+\left.\frac{U}{E}q(1+f_T)
  -\frac{U}{E}\frac{\psi_{01}}{FG}\right\}H'dr+\frac{3l}{8H^3},
\end{eqnarray}
whereas $X_{TF}$ takes the form as
\begin{eqnarray}\label{X}
  X_{TF} &=& -\frac{1}{2f_R}\left(\mu+\psi+\psi_{00}\right)+\frac{3}{2H^3}\int_0^r\frac{H^2}{f_R}\left\{
  \mu+\psi+\psi_{00}+\frac{U}{E}q(1+f_T)-\frac{U}{E}\frac{\psi_{01}}{FG}\right\}H'dr+\frac{3l}{8H^3}.
\end{eqnarray}
Thus, we are now able to construct a differential equation which shows a relationship between energy density inhomogeneity and weyl tensor.
\begin{eqnarray}
 \left( X_{TF}+\mu+\frac{1}{2f_{R}}(\mu+\psi+\psi_{00})\right)' &=&-3\frac{H'}{H}X_{TF}+\frac{(\Theta-\sigma)}{2f_R}\left(q(1+f_T)G+\frac{\psi_{01}}{G}\right),
\end{eqnarray}
If we choose $X_{TF}=0$ in the absence of dark source and dissipation, then we have
\begin{eqnarray}\label{t1}
  (\mu+\psi+\psi_{00})' &=& 0,
\end{eqnarray}
however, in general dissipative case it assumes the form
\begin{eqnarray}\label{t2}
  (\mu+\psi+\psi_{00})' &=& \frac{(\Theta-\sigma)}{2f_R}\left(q(1+f_T)G+\frac{\psi_{01}}{G}\right).
\end{eqnarray}
It shows that $X_{TF}$ controls the energy density homogeneity along with dark source terms.

\section{The Complexity Factor}

The definition of quantity measuring the complexity of a dynamical system is more generalized than for the static one as it faces two additional
factors. In static case, only fluid parameters are involved, while in non-static case, complexity of structure of system and of patterns
of evolution also contribute to the situation. For static case, definition is based on the assumptions that homogenous energy density and
isotropic pressure corresponds to the simplest system. However, for the later case, simplest possible patterns are also considered in order to
measure the degree of complexity of evolutionary patterns.

Recently, we have analyzed definition of complexity for anisotropic fluid non-static sphere in $f(R,T)$
gravity. We chose $Y_{TF}$ as complexity factor as it covers all the components that contributes to the complexity of a system.
In the case under consideration, we again found $Y_{TF}$ as most suitable scalar in order to analyze the components that trigger
complications in a system. It also incorporates the effects of dark source terms. We can see in Eq.(\ref{Y}) that it also contains the term
 comprising length of cylinder. Thus, it also measures the geometric variations in a system.

\section{The Homologous Evolution And The Homogeneous Expansion Condition}

After making the choice of $Y_{TF}$ as complexity factor, our next task is to analyze the complexity of evolutionary patterns
of the system. Such analysis involves two possibilities: the homologous condition and homogeneous expansion.
Homogeneous expansion corresponds to the zero value of prime derivative of expansion scalar which measures infinitesimal
changes in fluid distribution, whereas homologous evolution corresponds to the similarity of the patterns.

\subsection{The Homologous Evolution}
We can see that Eq.$(\ref{28})$ can be written as
\begin{eqnarray}
  D_H\left(\frac{U}{H}\right) &=& \frac{1+f_T}{f_R}\frac{q}{\tilde{E}}-\frac{1}{FG f_R \tilde{E}}\pi_{01}+\sqrt{3}\frac{\sigma}{H}.
\end{eqnarray}
whose integration leads to the equation
\begin{eqnarray}
  \frac{U}{H} &=&\int_0^r\left(\frac{1+f_T}{f_R}\frac{q}{\tilde{E}}-\frac{1}{FG f_R \tilde{E}}\pi_{01}+\sqrt{3}\frac{\sigma}{H}\right)H'dr+h(t),
\end{eqnarray}
where $h(t)$ is function of integration.
\begin{eqnarray}\label{101}
U &=&H\int_0^r\left(\frac{1+f_T}{f_R}\frac{q}{\tilde{E}}-\frac{1}{FG f_R \tilde{E}}\psi_{01}+\sqrt{3}\frac{\sigma}{H}\right)H'dr+Hh(t),
\end{eqnarray}
which yields
\begin{eqnarray}\label{102}
  U &=& \frac{U_\Sigma}{H_\Sigma}H-H\int_0^r\left(\frac{1+f_T}{f_R}\frac{q}{\tilde{E}}-\frac{1}{FG f_R \tilde{E}}\psi_{01}+\sqrt{3}\frac{\sigma}{H}\right)H'dr.
\end{eqnarray}
If the integral in Eq.$(\ref{101})$ and  Eq.$(\ref{102})$ vanishes, then $U\sim H$ which is characteristic of homologous evolution; it
would be possible if $\sigma=0$, $q=0$ and $\psi_{01}=0$ or the terms cancel each other.
\\
For homologous evolution, $U=h(t) H$ and $h(t)=\frac{U_\sigma}{H_\sigma}$ where $U=D_T H$. It makes us to follow that $H$ is separable and can be written as
\begin{eqnarray}\label{103}
  H &=& H_1(t)H_2(r)
\end{eqnarray}

The term with negative sign in Eq.$(\ref{102})$ shows that dissipation, shear and dark source entities are responsible
for the deviation of evolution from being homologous. Thus, we have
\begin{eqnarray}\label{105}
\frac{1+f_T}{f_R}\frac{qG}{H'}-\frac{1}{f_RF H'}\psi_{01}+\sqrt{3}\frac{\sigma}{H}  &=& 0
\end{eqnarray}
It represents homologous condition in general. For non-dissipative case, it takes the form as
\begin{eqnarray}\label{106}
\sqrt{3}\frac{\sigma}{H}  &=& \frac{1}{f_RF H'}\psi_{01}
\end{eqnarray}
It is obvious that homologous evolution does not correspond to the shear free condition in general, rather it depends on the choice
of $f(R,T)$ model.

\subsection{The homogeneous Expansion}
Homogeneous expansion also represents simple pattern of evolution.
Under homogeneous expansion, Eq.$(\ref{28})$ assumes the following form
\begin{eqnarray}\label{106i}
f_R\left(\sqrt{3} \frac{\sigma}{H}+\frac{1}{3}D_H\sigma\right)-\frac{1}{FH'}\psi_{01} &=& -\frac{qG}{H'}(1+f_T).
\end{eqnarray}
If we analyze the Eqs.$(\ref{105})$ and $(\ref{106i})$, then it can be clearly observe that imposition of these two conditions
 is followed by $D_H(\sigma)=0$. Here, its implication is based on the regularity conditions in the neighborhood of the center, that shear free
 condition implies zero dissipation.

\section{The $f(R, T)$ Model}
We can see that the result depends on the choice of $f(R,T)$ model. So we need to choose a viable $f(R, T)$ model in order to represent our results
in a meaningful way. The $f(R, T)$ model we have selected for discussion is developed by Sharif and Zubair \cite{zub} and has the following
mathematical form
\begin{eqnarray}\label{106a}
  f(R, T) &=& \alpha_1 R^m T^n +\alpha_2T(1+\alpha_3 T^p R^q),
\end{eqnarray}
where $\alpha_i's$ are positive real numbers , whereas $m, n, p, q$ assumes some value greater
than or equal to $1$ . We will analyze our results considering different cases of
above mentioned model and we will proceed our further discussion under following three cases:\\

\begin{enumerate}
  \item $f(R,T)= R+\alpha_2 T$, for $\alpha_1=1, m=1, n=0, \alpha_3=0$
  \item $f(R, T)= \alpha_1 R+\alpha_2 T+\alpha_4 T^2$, for $m=1, n=0, \alpha_4=\alpha_1\alpha_3, p=1, q=0$
  \item $f(R, T)= \alpha_1 R+\alpha_2 T(1+\alpha_3 TR^2)$, for $ m=1, n=0, p=1, q=2$
\end{enumerate}

\section{Kinematics and Dynamics of Stellar systems}

\subsection{Case I: $f(R,T)= R+\alpha_2 T$}

This form of model involves direct minimal curvature matter coupling, which has been used widely to explore
a number of cosmological phenomena because of its theoretical and cosmological consistency \cite{1*}.
 Many theoretical cosmological models of the universe have been proposed to analyze the behavior of
mysterious components and their physical and cosmological consequences are explored \cite{s6, s7}.
Galactic structures and their existence have also been discussed, and results are found in agreement with
previously established solutions and assumptions \cite{s8}.
For this model, homologous condition and homogeneous expansion condition given in Eqs.$(\ref{105})$ and $(\ref{106i})$ take the form as
\begin{eqnarray}\label{1071}
(1+\alpha_2)qG   &=&\sqrt{3}\frac{\sigma H'}{H},\\\label{1081}
\left(\sqrt{3} \frac{\sigma}{H}+\frac{1}{3}D_H\sigma\right)&=& -\frac{qG}{H'}(1+\alpha_2).
\end{eqnarray}
Here, Eq.$(\ref{1081})$ clearly depicts that fluid cannot be dissipative under shear free condition and
 homogeneous expansion. However, if Eqs.$(\ref{1071})$ and $(\ref{1081})$ hold at the same time, then we have
\begin{eqnarray}\label{1081a}
  (\Theta-\sqrt{3}\sigma)' &=& 0.
\end{eqnarray}
By inserting the values of expansion scalar and shear scalar given in Eqs.$(\ref{13})$ and $(\ref{sh})$, respectively,  we have
\begin{eqnarray}\label{1091}
 (\Theta-\sqrt{3}\sigma)' &=& \left(\frac{3}{F}\frac{\dot{H}}{H}\right)'= 0.
\end{eqnarray}
Eq.$(\ref{103})$ together with above equation follows that $F'=0$ which ensures the geodesic condition of the fluid. As $F$ possesses an arbitrary constant value, so we may choose $F=1$. With this choice, we get the following expression from Eq.(\ref{1091})
\begin{eqnarray}\label{1101}
\Theta-\sqrt{3}\sigma &=& 3\frac{\dot{H}}{H}.
\end{eqnarray}
We analyze this expression closer to the center and obtain the condition $(\Theta-\sqrt{3}\sigma)'=0$.
The successive derivatives of Eq.$(\ref{1101})$ with respect to $r$, also support our argument and strengthens the point that fluid is homologous.
\\
Again we analyze the Eq.$(\ref{28})$, if we
assume $\sigma=0$, (also fluid is non-dissipative), then we have
\begin{eqnarray}\label{1121}
  \Theta' &=&0.
\end{eqnarray}
It can clearly be observe that homologous patterns of the evolution imply homogeneity of the expansion scalar.
\\
If we assume $\Theta'=0$, then Eq.$(\ref{28})$ takes the following form
\begin{eqnarray}\label{1131}
 \left(\frac{\sqrt{3}\sigma}{H}-\frac{1}{\sqrt{3}}D_H(\sigma)\right) &=& 0,
\end{eqnarray}
which further takes the form as
\begin{eqnarray}\label{1131a}
\frac{\sigma'}{\sigma} &=& \frac{3R'}{R},
\end{eqnarray}
implying
\begin{eqnarray}\label{1131b}
\sigma &=& \frac{f_1(t)}{H^3},
\end{eqnarray}
where $f_1(t)$ is an arbitrary function of integration. As $r$ assumes zero value at the center, so $H$ will also be zero. Thus,
we must have $f_1(t)=0$ in order to avoid unboundedness of the expression. This situation lead to the vanishing of shear scalar. On the other hand,
if we choose the zero value for shear scalar, then  Eq.$(\ref{28})$ ensures the homogeneous expansion. Here, it is obvious that homogeneous expansion and
homologous condition imply each other in non dissipative case. Further, we have analyzed the situation
for homogeneous expansion in the presence of dissipation and obtained
\begin{eqnarray}\label{1131c}
  \sigma &=& -\frac{\sqrt{3}}{2H^3}\int^r_0 H^3 qG(1+\alpha_2)dr.
\end{eqnarray}
This expression makes it clear that homogeneous expansion and homologous condition are not compatible in the presence of dissipation.

Now, we consider some dynamical situations for the system under consideration.
As our previous discussion shows that fluid is
geodesic under  homologous condition in both dissipative and non-dissipative cases. Thus, if we apply
homologous condition on the Eq.(\ref{B4}), we obtain
\begin{eqnarray}\label{1141}
  D_TU &=& -\frac{m}{H^2}-\frac{H}{2}\left\{\alpha_2\mu -(1+\alpha_2)P_r+\frac{\alpha_2 T}{2}\right\}.
\end{eqnarray}
This equation can be re-written in the form of $Y_{TF}$ as
\begin{eqnarray}\label{1151}
 \frac{3 D_TU}{H} &=& -\frac{1}{2}\left\{(1-3\alpha_2)\mu-\alpha_2 T-2(1+\alpha_2)\Pi
 +3(1+\alpha_2)P_r\right\}+Y_{TF}-\frac{3l}{8H^3}.
\end{eqnarray}
Now, the manipulation of Eqs.$(\ref{16})$, $(\ref{18})$ and $(\ref{19})$ provides
\begin{eqnarray}\label{1151a}
 -\frac{2\ddot{H}}{H}-\frac{\ddot{G}}{G} &=& \frac{1}{2}\left\{(1-3\alpha_2)\mu-\alpha_2 T-2(1+\alpha_2)\Pi
 +3(1+\alpha_2)P_r\right\}+Y_{TF}-\frac{3l}{8H^3},
\end{eqnarray}
while the definition of velocity `U' of collapsing star provides
\begin{eqnarray}\label{1171}
  \frac{3 D_TU}{H}&=& \frac{3\ddot{H}}{H}.
\end{eqnarray}
Insertion of above two equations into Eq.$(\ref{1151a})$ leads to
\begin{eqnarray}\label{1181}
 Y_{TF} &=& \frac{\ddot{H}}{H}-\frac{\ddot{G}}{G}-\frac{3l}{8H^3}.
\end{eqnarray}
If  $Y_{TF}=0$, then Eq.$(\ref{1181})$ becomes
\begin{eqnarray}\label{118a}
\frac{3l}{8H^3} &=& \frac{\ddot{H}}{H}-\frac{\ddot{G}}{G}.
\end{eqnarray}
Since we are working on the assumption that fluid is homologous, so we can write the Eq.$(\ref{1151a})$ as
\begin{eqnarray}\label{1151b}
  3\left(\dot{h}(t)+h(t)\frac{\dot{H}}{H}\right) &=& -\frac{1}{2}\left\{(1-3\alpha_2)\mu-\alpha_2 T-2(1+\alpha_2)\Pi
 +3(1+\alpha_2)P_r\right\}+Y_{TF}-\frac{3l}{8H^3},
\end{eqnarray}
Now, we see the both cases when the fluid is non-dissipative or dissipative.

\subsubsection{The Dissipative and Non-dissipative Scenarios}

Here, we assume another condition that fluid is non-dissipative.
Under this assumption and for the choice of $f(R,T)$ model in case I, Eq.(\ref{106})
implies shear free condition for fluid configuration. With this implication, Eq.$(\ref{sh})$ leads to
\begin{eqnarray}\label{1251}
 \frac{\ddot{H}}{H}-\frac{\ddot{G}}{G}  &=& 0 \quad \Rightarrow\quad Y_{TF}=\frac{3l}{8H^3}.
\end{eqnarray}
Here, we can see that complexity factor $Y_{TF}=0$ is proportional to the fraction of $l$ and $H^3$.
We inserted this relation in Eq.$(\ref{Y})$ and found $\Pi=0$, which implies that $\mu'=0$.
Thus, $Y_{TF}=\frac{3l}{8H^3}$ represents
 simplest modes of evolution in the case of cylinder.
 Further, Eqs.$(\ref{118a})$ and $(\ref{1251})$
also strengthens our argument.

Now, we analyze the situation in the presence of dissipation. In this case, Eqs.$(\ref{1081})$ and $(\ref{sh})$ produce the following expression
\begin{eqnarray}\label{1261}
\dot{\sigma}  &=& \frac{1}{\sqrt{3}}\left\{\left(\frac{\dot{H}}{H}\right)^2- \left(\frac{\dot{G}}{G}\right)^2+\frac{3l}{8H^3}-Y_{TF}\right\}.
\end{eqnarray}
The time derivative of Eq.$(\ref{1081})$ and above equation provide the following mathematical expression
\begin{eqnarray}\label{1271}
  Y_{TF}\frac{H'}{H} &=& \frac{1}{2}Bq(1+\alpha_2)\left(\frac{\dot{q}}{q}+\frac{2\dot{B}}{B}+\frac{\dot{H}}{H}\right)+\frac{3l}{8H^3}.
\end{eqnarray}
If we assign zero value to complexity factor, then we get
\begin{eqnarray}\label{1281}
  Bq(1+\alpha_2)\left(\frac{\dot{q}}{q}+\frac{2\dot{B}}{B}+\frac{\dot{H}}{H}\right)+\frac{3l}{4H^3} &=& 0.
\end{eqnarray}
This differential equation can further be solved by using some suitable numerical or analytical methods of integration. It actually holds to represent
simplest dissipative regime.

\subsection{ Case-II: $f(R, T)= \alpha_1 R+\alpha_2 T+\alpha_4 T^2$}

This form of our selected model comprises linear and quadratic
terms in the trace of the energy-momentum tensor( EMT). The squared terms of EMT was first introduced in \cite{s9}.
This particular form of $f(R, T)$ model has been used to explore non-exotic matter wormholes \cite{s10}.
This type of choice usually contrast with higher order gravity and the results provide description of the universe
that enters from a decelerated phase of expansion  to an accelerated one and in agreement with observational data.
For this choice of $f(R,T)$ model, conditions obtained against simplest modes of evolution given in Eqs.$(\ref{105})$ and $(\ref{106i})$ take the form as
\begin{eqnarray}\label{129}
\frac{1+\alpha_2+2\alpha_4 T}{\alpha_1}qG   &=&\sqrt{3}\frac{\sigma H'}{H},\\\label{108a}
\left(\sqrt{3} \frac{\sigma}{H}+\frac{1}{3}D_H\sigma\right)&=& -\frac{qG}{\alpha_1H'}(1+\alpha_2+2\alpha_4 T)
\end{eqnarray}
Here, if we assume $\sigma=0$, Eq.(\ref{108a}) ensures the vanishing of dissipative variable. Thus, homogeneous expansion and
shear-free condition again ceases the fluid to be dissipative.
All the situations that exits in Eqs.(\ref{1131b}-\ref{108a}) are also valid in this case. However, in the presence of
dissipation, shear scalar assume the form as
\begin{eqnarray}\label{113*c}
  \sigma &=& -\frac{\sqrt{3}}{2H^3}\int^r_0 \frac{H^3}{\alpha_1} qG(1+\alpha_2+\alpha_4 T)dr.
\end{eqnarray}
In this case, dissipative variable again affects the homogeneous expansion and homologous condition and these are not found
compatible.

Now, we consider some dynamical situations for the system under consideration.
As our previous discussion shows that fluid is
geodesic under  homologous condition in both dissipative and non-dissipative cases. Thus, if we apply
homologous condition on the Eq.(\ref{B4}), we obtain
\begin{eqnarray}\label{114*}
  D_TU &=& -\frac{m}{H^2}-\frac{H}{2\alpha_1}\left\{(\alpha_2+2\alpha_4 T)\mu -(1+\alpha_2+2\alpha_4 T)P_r+\frac{(\alpha_2+2\alpha_4 T) T}{2}\right\}.
\end{eqnarray}
This equation can be re-written in the form of $Y_{TF}$ as
\begin{eqnarray}\nonumber
 \frac{3 D_TU}{H} &=& -\frac{1}{2\alpha_1}\left\{(1-3\alpha_2-6\alpha_4 T)\mu-\alpha_2 T-\alpha_4 T^2-2(1+\alpha_2+\alpha_4 T)\Pi
 +3(1+\alpha_2+\alpha_4 T)P_r\right\}\\\label{115*}&&+Y_{TF}-\frac{3l}{8H^3}.
\end{eqnarray}
Now, the manipulation of Eqs.$(\ref{16})$, $(\ref{18})$ and $(\ref{19})$ provides
\begin{eqnarray}\nonumber
 -\frac{2\ddot{H}}{H}-\frac{\ddot{G}}{G} &=& \frac{1}{2}\left\{(1-3\alpha_2-6\alpha_4 T)\mu-\alpha_2 T-2\alpha_4 T^2-2(1+\alpha_2+2\alpha_4 T)\Pi
 +3(1+\alpha_2 +2\alpha_4 T)P_r\right\}\\\label{115**}&&+Y_{TF}-\frac{3l}{8H^3},
\end{eqnarray}
while the definition of velocity `U' of collapsing star provides
\begin{eqnarray}\label{117*}
  \frac{3 D_TU}{H}&=& \frac{3\ddot{H}}{H},
\end{eqnarray}
Insertion of above two equations into Eq.$(\ref{115**})$ leads to
\begin{eqnarray}\label{118*}
 Y_{TF} &=& \frac{\ddot{H}}{H}-\frac{\ddot{G}}{G}-\frac{3l}{8H^3}.
\end{eqnarray}
If  $Y_{TF}=0$, then Eq.$(\ref{118*})$ becomes
\begin{eqnarray}\label{118a*}
\frac{3l}{8H^3} &=& \frac{\ddot{H}}{H}-\frac{\ddot{G}}{G}.
\end{eqnarray}
Since we are working on the assumption that fluid is homologous, so we can write the Eq.$(\ref{115**})$ as
\begin{eqnarray}\nonumber
  3\left(\dot{h}(t)+h(t)\frac{\dot{H}}{H}\right) &=& -\frac{1}{2}\left\{(1-3\alpha_2-6\alpha_4 T)\mu-\alpha_2 T-\alpha_4 T^2-2(1+\alpha_2+\alpha_4 T)\Pi
 +3(1+\alpha_2+\alpha_4 T)P_r\right\}+Y_{TF}\\\label{115w}&&-\frac{3l}{8H^3},
\end{eqnarray}

\subsubsection{The Dissipative and Non-dissipative Scenarios}

In non-dissipative case, we again observe the same scenario as it is discussed in case I. Thus, we need to discuss only dissipative case.
Here, we analyze the situation in the presence of dissipation. In this case, Eqs. $(\ref{sh})$ and $(\ref{108a})$ produce the following expression
\begin{eqnarray}\label{126p}
\dot{\sigma}  &=& \frac{1}{\sqrt{3}}\left\{\left(\frac{\dot{H}}{H}\right)^2- \left(\frac{\dot{G}}{G}\right)^2+\frac{3l}{8H^3}-Y_{TF}\right\}.
\end{eqnarray}
The time derivative of Eq.$(\ref{108a})$ and above equation provide the following mathematical expression
\begin{eqnarray}\label{127p}
  Y_{TF}\frac{H'}{H} &=& \frac{1}{2\alpha_1}Bq(1+\alpha_2+2\alpha_4 T)\left(\frac{\dot{q}}{q}
  +\frac{2\dot{B}}{B}+\frac{\dot{H}}{H}\right)+\frac{\alpha_4}{\alpha_2}\dot{T}+\frac{3l}{8H^3}.
\end{eqnarray}
If we assign zero value to complexity factor, then we get
\begin{eqnarray}\label{128p}
  \frac{Bq}{\alpha_1}(1+\alpha_2+2\alpha_4 T)\left(\frac{\dot{q}}{q}+\frac{2\dot{B}}{B}
  +\frac{\dot{H}}{H}\right)+2\frac{\alpha_4}{\alpha_2}\dot{T}+\frac{3l}{4H^3} &=& 0.
\end{eqnarray}
This differential equation can further be solved by using some suitable numerical or analytical methods of integration. It actually holds to represent simplest dissipative regime.

\subsection{ Case III: $f(R, T)= \alpha_1 R+\alpha_2 T(1+\alpha_3 TR^2)$}
This type of models offer the non-minimal coupling of curvature and matter components.
The similar type of choice has been recently used to measure the
impact of collision matter on the late-time
dynamics of $f(R, T)$ gravity \cite{s11}.
 In this case, homologous and
homogeneous conditions will take the form as
\begin{eqnarray}\label{lab1}
  \frac{1+\gamma_2}{\gamma_1}qG+\sqrt{3}\sigma \frac{H'}{H}&=& \frac{1}{\gamma_1}\left(\gamma_5-\frac{\dot{B}}{B}\gamma_4\right), \\\label{lab2}
  \gamma_1\left(\sqrt{3}\frac{\sigma}{H}+\frac{1}{\sqrt{3}}D_H\sigma\right)&=&\frac{1}{FH'}
 \left(\gamma_5-\frac{A'}{A}\gamma_3-\frac{\dot{B}}{B}\gamma_4\right),
\end{eqnarray}
however, Eq.(\ref{106}) takes the form as
\begin{eqnarray}\label{dq}
 \sqrt{3}\frac{\sigma}{H} &=&  \frac{1}{\gamma_1FH'}
 \left(\gamma_5-\frac{A'}{A}\gamma_3-\frac{\dot{B}}{B}\gamma_4\right),
\end{eqnarray}
where
\begin{eqnarray}\nonumber
\gamma_1&=&\alpha_1+2\alpha_4T^2 R,\\\nonumber
\gamma_2&=& \alpha_2+2\alpha_4TR^2,\\\nonumber
\gamma_3&=&\alpha_4(T^2\dot{R}+2 T \dot{T}R),\\\nonumber
\gamma_4&=&\alpha_4(T^2R'+2 T T'R),\\\nonumber
\gamma_5&=&2\alpha_4(T^2\dot{R'}+2TT'\dot{R}+2T'\dot{T}R+2T\dot{T'}R+2T\dot{T}R').
\end{eqnarray}
Here, Eq.(\ref{dq}) clearly shows that homologous evolution does not imply shear free condition in non-dissipative case.
Nevertheless, validity of the Eqs.$(\ref{lab1})$ and $(\ref{lab2})$ at the same time again makes us to believe that fluid is homologous.
However, if we choose $\Theta'=0$ and consider the Eq.$(\ref{28})$, then we have
\begin{eqnarray}\label{lab3}
  \sigma &=& \frac{\sqrt{3}}{2H^3}\int\frac{H^3}{F\gamma_1}\left(\gamma_5-\frac{F'}{F}\gamma_3
  -\frac{\dot{B}}{B}\gamma_4\right)dr.
\end{eqnarray}
Under the same condition (i.e., $q=0$), if we assume $\sigma=0$, then Eq.$(\ref{28})$ does not imply the homogeneous expansion, rather it takes the form as
\begin{eqnarray}\label{lab4}
  \Theta' &=& -\frac{\sqrt{3}}{F\gamma_1}\left(\gamma_5-\frac{F'}{F}\gamma_3-\frac{\dot{G}}{G}\gamma_4\right).
\end{eqnarray}
It is obvious from Eqs.$(\ref{lab3})$ and $(\ref{lab4})$ that homologous and homogeneous expansion conditions do not imply each other
in non-dissipative case. In the presence of dissipation, $\Theta'=0$ produce the following result
\begin{eqnarray}\label{lab5}
  \sigma &=& \frac{\sqrt{3}}{2H^3}\int\frac{H^3}{F\gamma_1}\left\{\left(\gamma_5-\frac{F'}{F}\gamma_3
  -\frac{\dot{B}}{B}\gamma_4\right)-FGq(1+\gamma_2)\right\}dr.
\end{eqnarray}
Again, we take into account dynamical considerations and apply homologous condition and obtain from \textcolor[rgb]{0.98,0.00,0.00}{\textbf{Eq.$(\ref{B4})$}}
\begin{eqnarray}\label{lab6}
  D_T U &=& -\frac{m}{H^2}-\frac{H}{2\gamma}\left\{\gamma_2 \mu+(1+\gamma_2)P_r+\phi+\phi_{11}\right\},
\end{eqnarray}
which further takes the form as
\begin{eqnarray}\label{lab7}
 \frac{3D_T U}{H} &=& -\frac{1}{2\gamma_1}\left\{(1+3\gamma_2)\mu-2\phi-2(1+\gamma_2)\Pi+\phi_{11}
 +2\phi_{22}+3(1+\gamma_2)P_r \right\}+Y_{TF}-\frac{3l}{8H^3},
\end{eqnarray}
where $\phi$ and $\phi_{ii}$ occur due to $f(R, T)$ extra degrees of freedom involved in the evolution.
The above expression can further be re-written as
\begin{eqnarray}\label{lab8}
 3\left(\dot{h}(t)+h(t)\frac{\dot{H}}{H}\right) &=& -\frac{1}{2\gamma_1}\left\{(1+3\gamma_2)\mu-2\phi-2(1+\gamma_2)\Pi+\phi_{11}
 +2\phi_{22}+3(1+\gamma_2)P_r \right\}+Y_{TF}-\frac{3l}{8H^3}.
\end{eqnarray}
Further, from field equations we have extracted the following result
\begin{eqnarray}\label{lab9}
 \frac{1}{2\gamma_1}\left\{(1+3\gamma_2)\mu-2\phi-2(1+\gamma_2)\Pi+\phi_{11}
 +2\phi_{22}+3(1+\gamma_2)P_r \right\}+Y_{TF}-\frac{3l}{8H^3}  &=& -\frac{2\ddot{H}}{H}-\frac{\ddot{G}}{G},
\end{eqnarray}
Using the definition of velocity `$U$' of collapsing star together with above equation, we again obtain the same result as given in Eq.(\ref{1181}) and vanishing of complexity factor $Y_{TF}$ yields the same expression as given in Eq.(\ref{118a}).

\subsubsection{The Dissipative and Non-dissipative Scenarios}

In this case, homologous condition does not imply shear free condition for non-dissipative scenario. Thus, under homologous condition,
we have $Y_{TF}= \frac{\ddot{H}}{H}-\frac{\ddot{G}}{G}-\frac{3l}{8H^3}$, which shows that complexity of the system is increased
in the presence of higher order curvature terms. Even in simplest modes of evolution, system have enough complexity index and fluid configuration does
not corresponds to the isotropic pressure and homogenous energy density.

Using Eqs.$(\ref{sh})$ and $(\ref{lab2})$, we find the relation for dissipative case as
\begin{eqnarray}\label{127pp}
  Y_{TF}\frac{H'}{H} &=& \frac{1}{2\alpha_1}Gq\frac{1+\gamma_2}{\gamma_1}\left(\frac{\dot{q}}{q}
  +\frac{2\dot{G}}{G}+\frac{\dot{H}}{H}\right)+\left(\frac{\gamma_2}{\gamma_1}\right)_{,0}qG-
  \frac{1}{\gamma_1}\left(\gamma_5-\frac{F'}{F}\gamma_3-\frac{\dot{G}}{G}\gamma_4\right)
  +\frac{3l}{8H^3}.
\end{eqnarray}
We can see dark source terms play important role and vanishing of complexity factor yields the following expression
\begin{eqnarray}\label{127ppp}
  \frac{1}{2\alpha_1}Gq\frac{1+\gamma_2}{\gamma_1}\left(\frac{\dot{q}}{q}
  +\frac{2\dot{G}}{G}+\frac{\dot{H}}{H}\right)+\left(\frac{\gamma_2}{\gamma_1}\right)_{,0}qG-
  \frac{1}{\gamma_1}\left(\gamma_5-\frac{F'}{F}\gamma_3-\frac{\dot{G}}{G}\gamma_4\right)
  +\frac{3l}{8H^3}&=& 0.
\end{eqnarray}

\section{Stability of The Vanishing Complexity Factor Condition}

In this section, our task is to find and analyze the conditions which are responsible
for an initial state of vanishing complexity factor under homologous condition. For this,
we need to develop the the evolution equation for structure scalar $Y_{TF}$ with the help
of Eqs.$(\ref{xx})$, $(\ref{yy})$, and (\ref{B2s}) which
takes the form as
\begin{eqnarray}\nonumber
  &&\dot{Y}_{TF}+(1+\alpha_2)\dot{\Pi}+\frac{3\dot{H}}{H}Y_{TF}
  +2(1+\alpha_2)\Pi\frac{\dot{H}}{H}+\frac{(1+\alpha_2)}{2}
  (\mu+P_r)\sigma-\frac{1+\alpha_2}{2}(\mu+P_r)\frac{\dot{G}}{G}
  -(1-\alpha_2^2)(\mu+P_\perp)\frac{\dot{H}}{H}\\\label{133*}&&+\alpha_2\frac{\dot{G}}{G}(\mu+P_\perp)-\frac{3}{2}\frac{H'}{H}(1+\alpha_2)q
  \frac{1-\alpha_2^2}{2G^2}(Gq)'+\frac{q(1-\alpha^2)}{G^2}\left(\frac{2H'}{H}-\frac{G'}{G}\right)+\alpha_2 q'+\Lambda= 0,
\end{eqnarray}
where $\Lambda$ contains dark source entities.
Now, we analyze this equation for dissipative and non-dissipative cases turn by turn. First we consider the non-dissipative case
at some initial moment where $Y_{TF}=q=\sigma=\Pi=0$, then previous equation takes the form as
\begin{eqnarray}\label{134*}
  \dot{Y}_{TF}+(1+\alpha_2)\dot{\Pi}-\frac{1-\alpha_2^2}{2}(\mu+P_\perp)\frac{\dot{H}}{H}+ \Lambda&=& 0.
\end{eqnarray}
In the most general case, when the system is dissipative, we have at the initial moment
\begin{eqnarray}\nonumber
  &&\dot{Y}_{TF}+(1+\alpha_2)\dot{\Pi}+2(1+\alpha_2)\Pi\frac{\dot{H}}{H}+\frac{1+\alpha_2}{2}
  (\mu+P_r)\sigma-\frac{1+\alpha_2}{2}(\mu+P_r)\frac{\dot{G}}{G}
  -(1-\alpha_2^2)(\mu+P_\perp)\frac{\dot{H}}{H}\\\label{135*}&&+\alpha_2\frac{\dot{G}}{G}(\mu+P_\perp)-\frac{3}{2}\frac{H'}{H}(1+\alpha_2)q
  \frac{1-\alpha_2^2}{2G^2}(Gq)'+\frac{q(1-\alpha^2)}{G^2}\left(\frac{2H'}{H}-\frac{G'}{G}\right)+\alpha_2 q'+ \Lambda= 0.
\end{eqnarray}
It is obvious from above equation that pressure anisotropy, energy density inhomogeneity, dissipative variable and dark source
entities all are crucial for complexity of a cylindrical system. Last term $\Lambda$ on the right hand side of the above equation
incorporates the effects of dark source terms which can be measured for any particular model.

\section{Conclusion}

The study of complexity on astrophysical scales is an interesting concept which helps the researchers to explore and identify
the factors that are responsible for the emergence of complexity in a system. In the pursuance of a complete and comprehensive definition of
complexity, Herrera \cite{12b} proposed a new definition for an anisotropic fluid sphere. Following Herrera's work, we have
explored the behavior of complexity factor for a cylindrically symmetric
dynamical object in $f(R,T)$ gravity. For this, we have explored field
equations for cylindrical symmetry and developed mass function using
 C-energy expression. We have constructed structure scalars for cylindrical geometry in $f(R,T)$ gravity and identified $Y_{TF}$ as complexity factor
among these scalars which incorporates the effects of
inhomogeneous energy density, anisotropic pressure and dissipative variable together with dark source terms.

The definition of complexity for a dynamical system involves not only the complexity of
the structure of the system, but also considers degree of complexity of the pattern
of evolution of the system. Thus, we have studied the complexity of patterns of evolution by assuming
two simplest modes of evolution,i.e., homologous evolution and homogeneous expansion.
We have worked out the conditions representing homogeneous expansion and homologous evolution, where we have
found that shear free condition implies zero dissipation if both of these conditions hold at the same time.
In order to
present detailed study in $f(R,T)$ gravity, we have selected a generic non-minimally coupled $f(R,T)$ model and discussed our findings for three
different cases of model under consideration.

\begin{itemize}
  \item In the first case, our $f(R,T)$ model is linear combination of first order terms of curvature `$R$' and trace of energy momentum tensor `$T$'.
Here, we have found that homogeneous expansion and shear free condition cease the fluid to be dissipative. We have also found that shear free
condition and homologous condition imply each other for this particular choice. In non-dissipative scenario, simplest modes of evolution are found to be
compatible with each other. However, in the presence of dissipative variable, they are not found compatible.
We have discussed dissipative and non-dissipative scenarios and found complexity factor is proportional to the fraction of
length of cylinder and curvature term of cubic order which represents minimum value of complexity factor when total amount of energy in
a cylindrical system is represented through gravitational C-energy.

\item In the second case, our model is of the form $f(R,T)=f_1(R)+f_2(T)$ , where  $f_1(R)$ is linear function, while $f_2(T)$ is
   quadratic one. All the results that are obtained in the first case are same for the second choice of $f(R,T)$ model. However in dissipative case,
  complexity of the system is increased due to the presence of quadratic $T$ terms.

   \item In the third case, our assumptions yields the form which consists on two parts. The first part coincides with
   case I, however second part contains product of quadratic terms of `$R$' and `$T$'. In this case, shear free
 condition and homologous condition are not found to be compatible for both dissipative and non-dissipative cases. It is observed that complexity
 of a system is increased in the presence of dark source entities, even in the simplest modes of evolution it does not attain the value that represents
 minimum complexity of the system.
 \end{itemize}
In the end, we analyzed the stability of vanishing complexity condition and observed different
outcomes in both dissipative and non-dissipative scenarios. In the absence of dissipation, significance of pressure isotropy and dark source entities is obvious from Eq.(\ref{134*}), however
in general scenario, Eq.(\ref{135*}) shows that multiple factors are involved.

We have compared our results with previous literature \cite{12a, s3} and found that
complexity factor for cylindrical system contains some extra terms because of
the geometrical difference of the self gravitating systems, which ceases it to attain zero value in
simplest modes of evolution. The complexity system for static self-gravitating systems has been explored
extensively, but it needs more study for dynamical systems. We intend to extend our work
for dynamical self-gravitating systems in the presence of charge.

\renewcommand{\theequation}{A.\arabic{equation}}
\setcounter{equation}{0}
\section*{Appendix A}

The divergence of the energy-momentum tensor is nonzero in $f(R, T)$ gravity and is found as
\begin{eqnarray}\setcounter{equation}{1}\label{B1s}
\nabla{^{\alpha}}T_{\alpha\beta}=\frac{f_T}{1-f_T}\left[(\Theta_{\alpha\beta}-T_{\alpha\beta})\nabla{^{\alpha}}ln
f_T-\frac{1}{2}g_{\alpha\beta}\nabla{^{\alpha}}T+\nabla^{^{\alpha}}\Theta_{\alpha\beta}\right].
\end{eqnarray}
Its divergence yields the following two equations:
\begin{eqnarray}\nonumber
&&\dot{\mu}\left(\frac{1-f_T-f_R f_T}{f_R(1-f_T)}\right)-\mu\frac{\dot{f_R}}{f_R^2}+
\frac{\dot{G}}{G}\frac{1}{f_R}(1+f_T)(\mu+P_r)+\frac{\dot{2H}}{H}\frac{1}{f_R}(1+f_T)(\mu+P_\perp)
+\left(\frac{\dot{T}}{2}\right)\frac{f_T}{1-f_T
}\\\nonumber&&-q'\frac{F}{G}\frac{(1+f_T)}{f_R}-\frac{q}{G^2}\left\{\frac{FG(1+f_T)}{f_R}\right\}_{,1}
-\frac{F}{G}\left(\frac{F'}{F}-\frac{G'}{G}+\frac{2H'}{H}\right)(1+f_T)q
+\left\{\frac{1}{f_R
}\left(\Psi                                                     +\psi_{00}\right)\right\}_{,0}\\\nonumber&&-\frac{1}{G^2}\left\{\frac{\psi_{01}}{f_R}\right\}_{,1}+{\frac{\psi_{01}}{G^{2}f_R}
}\left(\frac{F'}{F}-\frac{G'}{G}+\frac{2H'}{H}\right)+\frac{\psi_{00}}{f_R}\left(\frac{\dot{G}}{G
}+\frac{\dot{2H}}{H}\right)+\frac{\dot{G}}{G}\frac{\psi_{11}}{f_R}+\frac{\dot{2H}}{H}\frac{\psi_{22}}{f_R}
\\\label{B2s}&&-\left(\frac{2\dot{G}}{G}(\mu+P_r)+\frac{F}{G}2q'+4q\frac{F'}{G}\right)\frac{f_T}{1-f_T}
-\frac{F}{G}q\frac{f'_T}{1-f_T}=0,\\\nonumber
&&P'_r\left(\frac{1-f^{2}_T+2f_R
f_T}{f_R(1-f_T)}\right)+\frac{P_r}{f_R}\left(f'_T-\frac{f'_R(1+f_T)}{f_R}\right)+\mu'\frac{f_T}{f_R}+
\frac{\mu}{f_R}\left(f'_T-\frac{f_T f'_R}{f_R}\right)+\frac{F'}{F}\frac{(1+f_T)}{f_R}(\mu+P_r)\\\nonumber
&&+\frac{2H'}{H}\frac{1}{f_R}(1+f_T)(P_r-P_\perp)+\frac{f'_T}{1-f_T}(\mu+P_r)
-\frac{f_T}{1-f_T}\mu'+\frac{G}{F}\frac{(1+f_T)}{f_R}\dot{q}+\frac{q}{F^2}\left\{\frac{FG(1+f_T)}{f_R}\right\}_{,0}
\\\nonumber&&+
\left\{\frac{1}{f_R
}\left(\Psi+\psi_{11}\right)\right\}_{,1}-\frac{1}{F^2}\left\{\frac{\psi_{01}}{f_R}\right\}_{,0}+{\frac{\psi_{01}}{G^{2}f_R}}
\left(\frac{\dot{F}}{F}-\frac{\dot{G}}{G}+\frac{\dot{2H}}{H}\right)
+q\frac{G}{F}(1+f_T)\left(\frac{\dot{F}}{F}-\frac{\dot{G}}{G}+\frac{\dot{2H}}{H}\right)\\\nonumber&&+\frac{f_T}{1-f_T}\left(\frac{T'}{2}+\mu'\right)
+\frac{F'}{F}\frac{1}{f_R}\left(\psi_{00}+\pi_{11}\right)+\frac{2H'}{H}\frac{1}{f_R}\left(\pi_{11}-\psi_{22}\right)+\frac{2G}{F}q\left(1+\frac{F'}{F}\right)\frac{f_T}{1-f_T}
\\\label{B3s}&&+\frac{G}{F}\left(q\frac{\dot{f_T}}{f_R}+\frac{\dot{G}}{F}(\mu_+2P_r)\right)\frac{f_T}{1-f_T}
+\frac{f_T}{1-f_T}\frac{2}{A}(Bq)_{,0}
=0.
\end{eqnarray}
The acceleration $D_TU$ of a collapsing star can be obtained by manipulating the Eqs. $(\ref{13})$, $(\ref{17})$,  $(\ref{24})$ and  $(\ref{27})$:
\begin{eqnarray}\label{B4}
  D_TU &=& \frac{m}{H^2}-\frac{R}{2f_R}\left(1+f_T)P_r+f_T\mu+\pi+\pi_{11}\right)+Ea.
\end{eqnarray}

\vspace{.5cm}

\section*{Acknowledgments}

``Authors thank the Higher Education Commission, Islamabad, Pakistan for its
financial support under the NRPU project with grant number
$\text{5329/Federal/NRPU/R\&D/HEC/2016}$''.

\end{document}